\documentclass[twocolumn,amsmath,amssymb,pra,showpacs]{revtex4}

\usepackage{graphicx}

\begin{document}

\title{Atom Chip Diffraction of Bose-Einstein Condensates: \\The Role of Inter-Atomic Interactions}
\author{T.E. Judd}
\author{R.G. Scott}
\author{T.M. Fromhold}
\affiliation{School of Physics \& Astronomy, University of Nottingham, Nottingham NG7 2RD, United Kingdom}

\date{\today}

\begin{abstract}
We use supercomputer simulations to show that inter-atomic interactions can strongly affect the phase evolution of Bose-Einstein condensates that are diffracted from atom chips, thereby explaining recent experiments. Interactions broaden and depopulate non-central diffraction orders and so, counter-intuitively, they actually reduce the spatial width of the diffracted cloud. This means, more generally, that the experiments cannot be fully described by non-interacting theories and, consequently, interferometers that use many interacting atoms to enhance sensitivity require complex calibration. We identify device geometries required to approximate non-interacting behavior.
\end{abstract}

\pacs{03.75.Lm, 03.75.Dg, 37.25.+k}

\maketitle

\section{Introduction}
The ability to trap Bose-Einstein condensates (BECs) close to a surface using microchip technology provides a highly versatile way to manipulate and study the quantum wave nature of cold atoms \cite{andreas1,andreasdiff2,bongs,fernholz,amerongenyang,pasquini,reichelrev,hinds,dekker,folman,juddchaos,ottorig,hansel1,jonessurf,jointerf}. Consequently, atom chips form a clean and flexible environment for matter wave interferometry \cite{jointerf,wanginterf,schumminterf,jointerf2,halldwell}. Recent experiments \cite{andreas1,andreasdiff2} created an atom interferometer by reflecting a BEC from a magnetic grating fabricated on such a chip. Although evidence of successful diffraction was obtained, sharp, well-separated, diffraction orders were not observed. Furthermore, to extract phase information from the BEC, the experiments relied on two assumptions. Firstly, inter-atomic interactions do not change the BEC's phase in any way. Secondly, the reflection process is short enough to modulate only the phase of the BEC, but not its density. It is unclear to what extent these assumptions are valid and how they affect the accuracy of phase data inferred from the experiments.

In this paper, we show that inter-atomic interactions strongly affect the momentum distribution of the diffracted atom cloud, and hence its phase evolution, by broadening the diffraction orders and depopulating the higher ones.  In addition, the reflection process is long enough to modulate strongly both the phase {\em{and}} density of the condensate. When these effects are included, our calculations give good quantitative agreement with the diffraction patterns observed in the experiments. Interferometers based on interacting particles could make sensitive measurement devices \cite{scottjuddinterf,jonessurf,jointerf2}, but our results show that they respond in a highly complex way to spatially-varying phase patterns. To recover the essentially non-interacting behavior required for simple analytical models to make accurate {\textit{a priori}} predictions of an interferometer's characteristics, we find that the kinetic energy of the BEC's separating diffraction orders must exceed the peak inter-atomic interaction energy.

The diffraction process has two distinct stages: the imprinting stage in which the BEC reflects from the grating, and the expansion stage in which the atom cloud is released from its trap \cite{andreas1,andreasdiff2}. The imprinting process encodes a diffraction pattern in momentum space, which, after time-of-flight expansion, manifests itself in a real-space diffraction pattern. Our goal is to understand the role of interactions and excitations at each stage. We therefore perform separate simulations for the reflection-imprinting process and the subsequent expansion.

\section{System and Methodology}

\begin{figure}
\includegraphics[width=1.0\columnwidth]{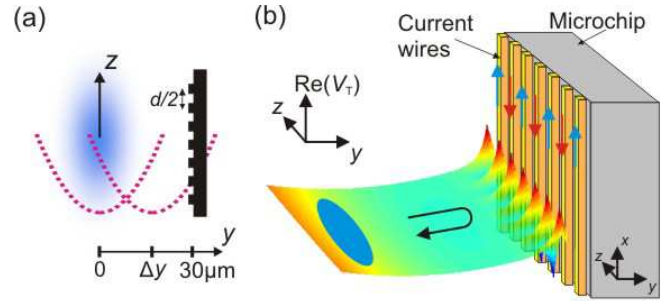} 
\caption{\label{fig:fullsys}(Color) (a) Schematic cross-section (black shape) in the $x=0$ plane (axes inset) through the microchip and wires, which are separated by $d/2$. Blue shape shows the BEC in its initial position centered at $y=0$. Left (right) dotted parabola shows the harmonic trap potential before (after) displacement $\Delta y$. (b) Curved surface: real part of the potential energy $V_T(y,z)$ (defined in the text, axes inset) after trap displacement. The displaced trap accelerates the BEC (blue) towards the microchip and wire grating on the right (axes inset). Vertical arrows on the wires indicate direction of current. Curved black arrow indicates the BEC's motion path. The offset magnetic field is not shown for clarity.}
\end{figure}

Figure \ref{fig:fullsys} shows a schematic diagram of the system used to imprint the BEC. 
In the experiments \cite{andreas1,andreasdiff2} and our calculations, a BEC of 1.2$\times$10$^5$ $^{87}$Rb atoms is prepared in a harmonic trap with frequencies $\omega_z=2\pi\times$16 rad s$^{-1}$ and $\omega_r=2\pi\times$76 rad s$^{-1}$ in the longitudinal ($z$) and radial ($x$-$y$ plane) directions respectively. The peak atom density ($n_0\approx 10^{20}$\:m$^{-3}$) is initially at $x=y=z=0$. An array of current carrying wires is microfabricated on a Si surface that lies in the $y=30\:\mu$m plane [Fig.\ \ref{fig:fullsys}(a)]. Adjacent wires are separated by 2\:$\mu$m [Fig.\ \ref{fig:fullsys}(a)] and carry a current of 0.2\:mA in alternating directions [Fig.\ \ref{fig:fullsys}(b)], giving a grating period $d=4$\:$\mu$m. An offset field of 2\:G in the $z$-direction combines with the field from the grating wires to create a potential profile that oscillates strongly between peaks and troughs, thus forming a diffraction grating [Fig.\ \ref{fig:fullsys}(b)] within $\sim$5\:$\mu$m of the surface \cite{footsurf}. 

At time $t=0$, we displace the harmonic trap through a distance $\Delta y = 11$ -- $15$\:$\mu$m in the $y$-direction and hence accelerate the BEC towards the grating along its short radial axis [Fig.\ \ref{fig:fullsys}(a)]. Since $\Delta y < 30$\:$\mu$m, the BEC must rise up the other side of the trap and decelerate before making contact with the grating potential. Impact velocities, $v_y$ (determined by $\Delta y$), are in the range $0$ -- $4$\:mm s$^{-1}$. The imprinting stage lasts 12\:ms, by which time the condensate has approximately returned to its original position.

We determine the dynamics of the BEC in 3D by solving the Gross-Pitaevskii equation (GPE)
\begin{equation}
\label{eq:gpe}
-\frac{\hbar^2}{2m}\nabla^2\Psi+V_{T}\Psi+\frac{4\pi\hbar^{2}a}{m}\left|\Psi\right|^2\Psi=i\hbar\frac{\partial\Psi}{\partial t},
\end{equation}
where $m$ is the mass of a single $^{87}$Rb atom, $a=5.4$\:nm is the $s$-wave scattering length, $\nabla^2$ is the Laplacian, $V_T(y,z)$ is the potential energy produced by the magnetic trap, grating, and chip surface [Fig.\ \ref{fig:fullsys}(b)], and $\Psi(x,y,z)$ is the wave function at time $t$, normalized so that $\left|\Psi\right|^2$ is the number of atoms per unit volume. We solve the equation using the Crank-Nicolson \cite{scott} method, and also use the Truncated Wigner \cite{steelTW} approach to investigate the effects of quantum fluctuations. Since the potential landscape contains very sharp features [Fig.\ \ref{fig:fullsys}(b)], solving Eq. \ref{eq:gpe} requires parallel programming implemented on a supercomputer to achieve the required resolution \cite{footres}.

\section{Results and Discussion}

\subsection{The imprinting process}

\begin{figure}
\includegraphics[width=1.0\columnwidth]{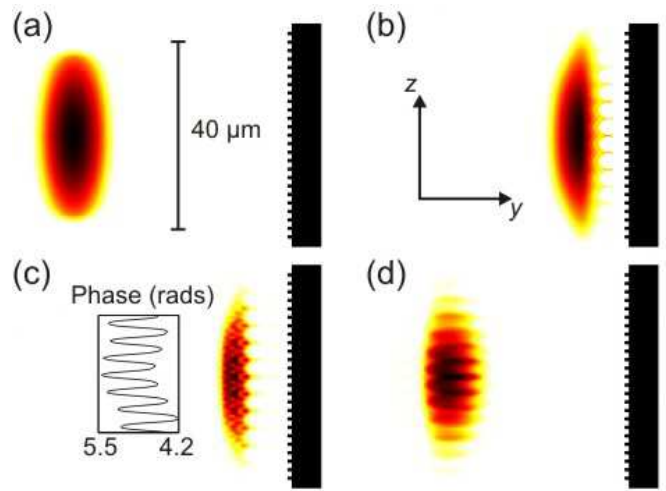} 
\caption{\label{fig:snaps}(Color online) Density slices (dark high) in the $x=0$ plane [axes in (b)] of the diffracting BEC  at $t=$ (a) 0.0\:ms, (b) 5.6\:ms, (c) 7.6\:ms and (d) 12.0\:ms. Bar in (a) shows scale. Black shapes (not to scale) indicate the position of the atom chip and grating. Inset in (c) shows phase imprint along $z$ (vertical) at $t = 7.6$\:ms.}
\end{figure}

Figure \ref{fig:snaps} shows density slices of the BEC in the $x=0$ plane as it reflects from the wire array following a trap displacement $\Delta y = 13$\:$\mu$m. 
Figure \ref{fig:snaps}(a) shows the initial state at $t = 0$\:ms. Figure \ref{fig:snaps}(b) shows the BEC just after contact with the lattice potential at $t = 5.6$\:ms and Fig.\ \ref{fig:snaps}(c) shows the BEC as it begins to move away from the lattice at $t = 7.6$\:ms. The atoms interact strongly with the lattice for a time interval $\Delta t\simeq 1$ -- $2$\:ms, which depends on $v_y$. By the end of this time interval, a density imprint and a clear phase imprint \cite{footimp} have formed [Fig.\ \ref{fig:snaps}(c) plus inset]. At $t = 12.0$\:ms, just before the trap is switched off, the density imprint is still evident [Fig.\ \ref{fig:snaps}(d)]. For all trap displacements, our calculations reveal phase and density imprints that vary approximately sinusoidally across the reflected BEC. 

Previous work predicts that BEC reflection processes can produce vortices in certain parameter regimes \cite{scott,scottol,scottari}. Here, however, no vortices form for any trap displacements because $\Delta t$ is always shorter than the time required for sound waves to propagate to the edge of the condensate and thus trigger the formation of excitations \cite{scott}. In addition, quantum fluctuations have been shown to produce decoherence and scattering halos in reflected BECs. For the parameters considered here, though, they make no significant difference to the BEC's dynamics because the velocities involved are too low to generate significant incoherent scattering \cite{scottnoise}. The bare GPE therefore captures the key physics and so we use it to produce all the results presented here.

\subsection{Broadening and depopulation of non-central diffraction orders}

We now perform a detailed analysis of the role of interactions in the diffraction process. To isolate their effects, we consider a simple model, which does not depend on the details of the imprinting process, only on the BEC's mean-field response to a sinusoidal phase imprint $\phi=S$\:sin$(2\pi z/d)$, like that produced by the grating. We use $S=1.4$ for all our expansion simulations as this is a typical value deduced from experiment \cite{andreas1,andreasdiff2}.

\begin{figure}
\includegraphics[width=1.0\columnwidth]{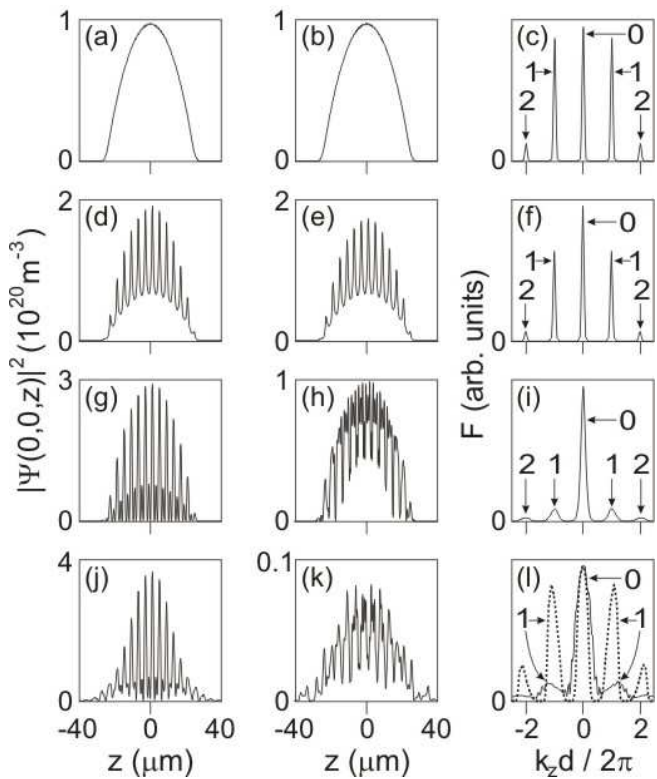} 
\caption{\label{fig:diffpopns}Left (center) column: density profiles along the $z$-axis for a non-interacting atom cloud (interacting BEC) after a sinusoidal phase imprint. Right column: Fourier power spectra calculated for the interacting BEC. The first, second, third and fourth rows down show, respectively, profiles at $0$\:ms, 0.2\:ms, 1\:ms, and 8\:ms after the imprint. Dashed curve in (l) shows the momentum distribution for diffraction from a grating with half the original period (see text).}
\end{figure}

Figure \ref{fig:diffpopns} shows how the density profile along $z$ evolves after phase imprinting a non-interacting atom cloud (left-hand column) and an interacting BEC (central column). The initial density profiles without and with interactions [Fig.\ \ref{fig:diffpopns}(a) and \ref{fig:diffpopns}(b) respectively] are taken to be identical to simplify the comparison of their subsequent evolution. The third column in Fig.\ \ref{fig:diffpopns} shows $k$-space momentum distributions for the interacting BEC, obtained by integrating the wavefunction along the $x$- and $y$-directions and calculating the one-dimensional Fourier power spectrum, $F$, along the $z$-direction. The initial momentum distribution has sharp peaks [Fig.\ \ref{fig:diffpopns}(c), arrowed] at the 0th, 1st and 2nd diffraction orders.

The second row of Fig.\ \ref{fig:diffpopns} shows the density profiles and momentum distribution 0.2\:ms after the phase imprinting. The density profiles for the non-interacting [Fig.\ \ref{fig:diffpopns}(d)] and interacting case [Fig.\ \ref{fig:diffpopns}(e)] now contain sharp modulations, which are interference fringes between emerging diffraction orders. Both density profiles look very similar and the momentum distribution for the interacting case [Fig.\ \ref{fig:diffpopns}(f)] has barely changed from its initial form [Fig.\ \ref{fig:diffpopns}(c)], indicating that the interactions have a negligible effect on this timescale. This is because 0.2\:ms is comparable with the minimum correlation time, $t_c = 0.14$\:ms \cite{foottc}, which means that collective interaction effects have had insufficient time to develop.

Figure \ref{fig:diffpopns} reveals that we cannot, as in previous analyses \cite{andreas1,andreasdiff2}, assume that there is no change in the atom density profile during the imprinting stage. Since $\Delta t \simeq 1$ -- $2$\:ms, it is not meaningful to speak of a pure phase imprint in the experiments \cite{andreas1} because a combination of phase and density modulations inevitably occurs on this timescale. It may, however, be possible to quantify a heuristic ``overall imprint'' strength. To obtain a pure phase imprint requires $\Delta t \lesssim 0.1$\:ms. This could be achieved by flashing the grating on briefly, by turning the current on and off rapidly, as opposed to having it on permanently.

The third row of Fig.\ \ref{fig:diffpopns} shows the density and momentum profiles $1$\:ms after imprinting. Differences between the non-interacting and interacting cases now become apparent because the system has evolved for several correlation times. In the non-interacting profile [Fig.\ \ref{fig:diffpopns}(g)], the density minima extend all the way to zero whereas they generally do not in the interacting case [Fig.\ \ref{fig:diffpopns}(h)] \cite{foot3rdorder}. The amplitude of the density modulation is also much lower in the interacting case [note the \textit{change of vertical scale} between Figs. \ref{fig:diffpopns}(g) and \ref{fig:diffpopns}(h)]. The momentum distribution of the BEC [Fig.\ \ref{fig:diffpopns}(i)] reveals dramatic depopulation of the 1st and 2nd order diffraction peaks [arrowed]. Counter-intuitively, the repulsive interactions therefore {\it{reduce}} the width of the BEC in momentum space. This demonstrates that the experimental data cannot be fully analysed using a non-interacting theory. Such theories assume that the order populations do not change during diffraction; this is clearly violated in the interacting case.

Depopulation occurs in the interacting case because mean-field repulsion pushes atoms away from high density regions into the gaps between the peaks. This prevents the density minima from falling close to zero [Fig.\ \ref{fig:diffpopns}(h)] and broadens the modulations, thereby reducing the number of atoms in the higher momentum diffraction orders [Fig.\ \ref{fig:diffpopns}(i)]. In the non-interacting case, there is no mean-field resistance to sharp modulation of the density profile and the minima fall close to zero [Fig.\ \ref{fig:diffpopns}(g)]. Interestingly, interactions do not significantly change the positions of the diffraction orders along the $k_z$ axis [compare Figs. \ref{fig:diffpopns}(c) and \ref{fig:diffpopns}(i)], and hence the speeds at which distinct diffraction orders separate are largely unaffected.

The fourth row of Fig.\ \ref{fig:diffpopns} shows the density and momentum profiles at $8$\:ms after imprinting [solid curves]. The non-interacting density profile [Fig.\ \ref{fig:diffpopns}(j)] has retained its sharp modulations, which extend down to zero. In the interacting case [Fig.\ \ref{fig:diffpopns}(k)], however, the density modulations have become even broader and lower in amplitude. At this point in the expansion process, the density of the BEC has become so low that interactions no longer play a significant role, meaning that the diffraction peaks subsequently separate at a speed determined simply by their positions in momentum space. Consequently, the corresponding momentum distribution [solid curve in Fig.\ \ref{fig:diffpopns}(l)] has the same shape as the far-field spatial density profile. It reveals further depopulation of the higher diffraction orders and broadening of all the peaks. This shows that the experiments could never observe spatially well-separated diffraction peaks, even for very long expansion times ($> 100$\:ms). Quantum fluctuations were again found to make no significant difference to these simulations, with coherence losses always $\sim 1\%$.

\subsection{Comparison with experiment}

\begin{figure}
\includegraphics[width=1.0\columnwidth]{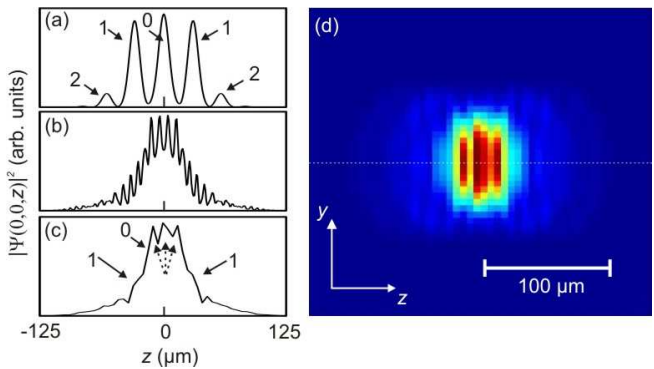} 
\caption{\label{fig:exp}(Color) (a)-(c) Density profiles along the $z$-axis after a sinusoidal phase imprint and 26\:ms expansion time  for the non-interacting (a) and interacting (b) clouds. (c) Curve obtained after passing the trace in (b) through a low resolution filter. (d) Density profile of the interacting cloud in the $y$-$z$ plane (axes inset), obtained by integrating $|\Psi(x,y,z)|^2$ along $x$ after a sinusoidal imprint, 26\:ms expansion and application of a low pass filter to simulate experimental image resolution. Horizontal bar shows scale. Density profile in (c) is taken along dotted line in (d).}
\end{figure}

To compare our theory with the diffracted BEC density profiles seen in the experiments \cite{andreasdiff2}, we now perform longer expansion simulations, which approximate the experimental procedure. After imprinting both phase and density profiles similar to those shown in Fig.\ \ref{fig:snaps}(c), the BEC is held in the harmonic trap for 6\:ms and then the trap is turned off, allowing the BEC to expand freely for 20\:ms.
Figures \ref{fig:exp}(a) and \ref{fig:exp}(b) show density profiles along the $z$-axis at the end of this process for a non-interacting matter wave (a) and a BEC (b). 
For the non-interacting atom cloud [Fig.\ \ref{fig:exp}(a)], the diffraction peaks separate and a typical far-field diffraction pattern appears. The $0$th, 1st and 2nd diffraction orders (arrowed) are all significantly populated.
For the interacting BEC [Fig.\ \ref{fig:exp}(b)], 20 ms expansion is insufficient to reach the far-field regime. Hence the density profile has rapid modulations along the $z$-direction, which are interference fringes between different diffraction orders. As explained in our discussion of Fig.\ \ref{fig:diffpopns}(h), the amplitude of the fringes has been reduced by the inter-atomic interactions. Figure \ref{fig:exp}(b) also shows that the spread of the cloud is significantly less than the non-interacting BEC. This emphasises the counter-intuitive effect of repulsive interactions in a diffraction system and suggests that attempts to infer phase data from atom interferometers based on BECs may give rise to major systematic errors if interactions are not properly accounted for.

Figure \ref{fig:exp}(c) shows the effect of passing the curve in Fig.\ \ref{fig:exp}(b) through a low-pass filter with the same resolution as the camera used to image the BECs in experiment (pixel size $\sim 5$\:$\mu$m). Figure \ref{fig:exp}(d) shows the corresponding (filtered) 2D density profile in the $y$-$z$ plane after integrating $|\Psi(x,y,z)|^2$ along $x$. We now obtain good quantitative agreement with the experimental data [compare Fig.\ \ref{fig:exp}(d) with Fig.\ 2(a), right-hand panel, of Ref.\ \cite{andreasdiff2}]. Our simulated density profile has approximately the same shape and width as the experimental profile, and both contain three small peaks [dotted arrows in Fig.\ \ref{fig:exp}(c)], corresponding to aliased interference fringes between separating diffraction orders. As in Fig.\ \ref{fig:diffpopns}, these peaks are dramatically weakened compared with the non-interacting case where nodes are observed [Fig.\ \ref{fig:diffpopns}(g)]. We also reproduce the ``shoulders'' [labelled ``1'' in Fig.\ \ref{fig:exp}(c)] observed in experiment, which are remnants of the 1st diffraction orders.
Since our zero-temperature mean-field theory agrees well with experiment, we infer that possible deleterious effects due to heating of the atom cloud during diffraction are not significant.

\subsection{Achieving a non-interacting regime}

We now consider the parameter regimes required to approximate non-interacting behavior. We find that we approach this regime gradually as the kinetic energy associated with the first diffraction order, $E_D = 2\pi^2\hbar^2/m d^2$ \cite{wimbergerOL}, becomes comparable to, or greater than, the BEC's peak interaction energy $E_I = 4\pi\hbar^2an_0/m$. This condition can be satisfied by halving the original grating period to $d = 2$\:$\mu$m. In this case, after 8\:ms expansion, the diffraction peaks [dashed curve in Fig.\ \ref{fig:diffpopns}(l)] are very similar to the non-interacting case [Fig.\ \ref{fig:diffpopns}(c)] and reveal far less depopulation and broadening than when $d = 4$\:$\mu$m [solid curve in Fig.\ \ref{fig:diffpopns}(l)]. We obtain similar results by keeping $d = 4$\:$\mu$m but reducing the BEC's density so that the condition $E_D \gtrsim E_I$ is satisfied. An alternative approach is to increase radial confinement so that the BEC expansion occurs on a faster timescale than the depopulation of the diffraction orders.

\section{Conclusion}

In summary, we have studied how inter-atomic interactions affect the dynamics of BECs diffracting from magnetic gratings on atom chips. Analysis of our numerical calculations reveals that if $E_I \gtrsim E_D$, as in recent experiments \cite{andreas1,andreasdiff2}, inter-atomic interactions significantly broaden the diffraction orders and depopulate the higher ones. This means that the BEC's phase evolution cannot be described accurately by simple non-interacting theories, which omit key physical processes and therefore lead to qualitatively incorrect predictions. Atom chips like the grating considered here, which imprint well-defined patterns in the phase and density profiles of a BEC, are commonly used as atom interferometers. Our results show that inter-atomic interactions can, generically, limit an atom interferometer's sensitivity to phase imprints. Although such interferometers can function by exploiting inter-atomic interactions \cite{jointerf2,scottjuddinterf}, their adverse effect on device sensitivity and accuracy can be reduced by operating in a regime where simple non-interacting theories can reliably predict the device characteristics.

\begin{acknowledgments}
We thank A. G\"unther, C. Zimmermann and J. Fort\'agh for helpful discussions. This work was funded by EPSRC. 
\end{acknowledgments}

\bibliography{biblio}

\begin{thebibliography}{31}
\expandafter\ifx\csname natexlab\endcsname\relax\def\natexlab#1{#1}\fi
\expandafter\ifx\csname bibnamefont\endcsname\relax
  \def\bibnamefont#1{#1}\fi
\expandafter\ifx\csname bibfnamefont\endcsname\relax
  \def\bibfnamefont#1{#1}\fi
\expandafter\ifx\csname citenamefont\endcsname\relax
  \def\citenamefont#1{#1}\fi
\expandafter\ifx\csname url\endcsname\relax
  \def\url#1{\texttt{#1}}\fi
\expandafter\ifx\csname urlprefix\endcsname\relax\def\urlprefix{URL }\fi
\providecommand{\bibinfo}[2]{#2}
\providecommand{\eprint}[2][]{\url{#2}}

\bibitem[{\citenamefont{G{\"u}nther\textit{ et al.}}(2005)}]{andreas1}
\bibinfo{author}{\bibfnamefont{A.}~\bibnamefont{G{\"u}nther\textit{ et al.}}},
  \bibinfo{journal}{Phys. Rev. Lett.} \textbf{\bibinfo{volume}{95}},
  \bibinfo{pages}{170405} (\bibinfo{year}{2005}).

\bibitem[{\citenamefont{G{\"u}nther et~al.}(2007)\citenamefont{G{\"u}nther,
  Kraft, Zimmermann, and Fort{\'{a}}gh}}]{andreasdiff2}
\bibinfo{author}{\bibfnamefont{A.}~\bibnamefont{G{\"u}nther}},
  \bibinfo{author}{\bibfnamefont{S.}~\bibnamefont{Kraft}},
  \bibinfo{author}{\bibfnamefont{C.}~\bibnamefont{Zimmermann}},
  \bibnamefont{and}
  \bibinfo{author}{\bibfnamefont{J.}~\bibnamefont{Fort{\'{a}}gh}},
  \bibinfo{journal}{Phys. Rev. Lett.} \textbf{\bibinfo{volume}{98}},
  \bibinfo{pages}{140403} (\bibinfo{year}{2007}).

\bibitem[{\citenamefont{Bongs\textit{ et al.}}(1999)}]{bongs}
\bibinfo{author}{\bibfnamefont{K.}~\bibnamefont{Bongs\textit{ et al.}}},
  \bibinfo{journal}{Phys. Rev. Lett.} \textbf{\bibinfo{volume}{83}},
  \bibinfo{pages}{3577} (\bibinfo{year}{1999}).

\bibitem[{\citenamefont{Fernholz et~al.}(2008)\citenamefont{Fernholz,
  Gerritsma, Whitlock, Barb, and Spreeuw}}]{fernholz}
\bibinfo{author}{\bibfnamefont{T.}~\bibnamefont{Fernholz}},
  \bibinfo{author}{\bibfnamefont{R.}~\bibnamefont{Gerritsma}},
  \bibinfo{author}{\bibfnamefont{S.}~\bibnamefont{Whitlock}},
  \bibinfo{author}{\bibfnamefont{I.}~\bibnamefont{Barb}}, \bibnamefont{and}
  \bibinfo{author}{\bibfnamefont{R.~J.~C.} \bibnamefont{Spreeuw}},
  \bibinfo{journal}{Phys. Rev. A} \textbf{\bibinfo{volume}{77}},
  \bibinfo{pages}{033409} (\bibinfo{year}{2008}).

\bibitem[{\citenamefont{van Amerongen et~al.}(2008)\citenamefont{van Amerongen,
  van Es, Wicke, Kheruntsyan, and van Druten}}]{amerongenyang}
\bibinfo{author}{\bibfnamefont{A.~H.} \bibnamefont{van Amerongen}},
  \bibinfo{author}{\bibfnamefont{J.~J.~P.} \bibnamefont{van Es}},
  \bibinfo{author}{\bibfnamefont{P.}~\bibnamefont{Wicke}},
  \bibinfo{author}{\bibfnamefont{K.~V.} \bibnamefont{Kheruntsyan}},
  \bibnamefont{and} \bibinfo{author}{\bibfnamefont{N.~J.} \bibnamefont{van
  Druten}}, \bibinfo{journal}{Phys. Rev. Lett.} \textbf{\bibinfo{volume}{100}},
  \bibinfo{pages}{090402} (\bibinfo{year}{2008}).

\bibitem[{\citenamefont{Pasquini\textit{ et al.}}(2004)}]{pasquini}
\bibinfo{author}{\bibfnamefont{T.~A.} \bibnamefont{Pasquini\textit{ et al.}}},
  \bibinfo{journal}{Phys. Rev. Lett} \textbf{\bibinfo{volume}{93}},
  \bibinfo{pages}{223201} (\bibinfo{year}{2004}).

\bibitem[{\citenamefont{Reichel}(2002)}]{reichelrev}
\bibinfo{author}{\bibfnamefont{J.}~\bibnamefont{Reichel}},
  \bibinfo{journal}{Appl. Phys. B} \textbf{\bibinfo{volume}{75}},
  \bibinfo{pages}{469} (\bibinfo{year}{2002}).

\bibitem[{\citenamefont{Hinds and Hughes}(1999)}]{hinds}
\bibinfo{author}{\bibfnamefont{E.~A.} \bibnamefont{Hinds}} \bibnamefont{and}
  \bibinfo{author}{\bibfnamefont{I.~G.} \bibnamefont{Hughes}},
  \bibinfo{journal}{J. Phys. D: Appl. Phys.} \textbf{\bibinfo{volume}{32}},
  \bibinfo{pages}{R119} (\bibinfo{year}{1999}).

\bibitem[{\citenamefont{Dekker\textit{ et al.}}(2000)}]{dekker}
\bibinfo{author}{\bibfnamefont{N.~H.} \bibnamefont{Dekker\textit{ et al.}}},
  \bibinfo{journal}{Phys. Rev. Lett.} \textbf{\bibinfo{volume}{84}},
  \bibinfo{pages}{1124} (\bibinfo{year}{2000}).

\bibitem[{\citenamefont{Folman\textit{ et al.}}(2000)}]{folman}
\bibinfo{author}{\bibfnamefont{R.}~\bibnamefont{Folman\textit{ et al.}}},
  \bibinfo{journal}{Phys. Rev. Lett.} \textbf{\bibinfo{volume}{84}},
  \bibinfo{pages}{4749} (\bibinfo{year}{2000}).

\bibitem[{\citenamefont{Judd\textit{ et al.}}(2007)}]{juddchaos}
\bibinfo{author}{\bibfnamefont{T.~E.} \bibnamefont{Judd\textit{ et al.}}},
  \bibinfo{journal}{Prog. Theo. Phys. Supp.} \textbf{\bibinfo{volume}{166}},
  \bibinfo{pages}{169} (\bibinfo{year}{2007}).

\bibitem[{\citenamefont{Ott et~al.}(2001)\citenamefont{Ott, Fort{\'{a}}gh,
  Schlotterbeck, Grossmann, and Zimmermann}}]{ottorig}
\bibinfo{author}{\bibfnamefont{H.}~\bibnamefont{Ott}},
  \bibinfo{author}{\bibfnamefont{J.}~\bibnamefont{Fort{\'{a}}gh}},
  \bibinfo{author}{\bibfnamefont{G.}~\bibnamefont{Schlotterbeck}},
  \bibinfo{author}{\bibfnamefont{A.}~\bibnamefont{Grossmann}},
  \bibnamefont{and}
  \bibinfo{author}{\bibfnamefont{C.}~\bibnamefont{Zimmermann}},
  \bibinfo{journal}{Phys. Rev. Lett.} \textbf{\bibinfo{volume}{87}},
  \bibinfo{pages}{230401} (\bibinfo{year}{2001}).

\bibitem[{\citenamefont{H{\"a}nsel\textit{ et al.}}(2001)}]{hansel1}
\bibinfo{author}{\bibfnamefont{W.}~\bibnamefont{H{\"a}nsel\textit{ et al.}}},
  \bibinfo{journal}{Nature} \textbf{\bibinfo{volume}{413}},
  \bibinfo{pages}{498} (\bibinfo{year}{2001}).

\bibitem[{\citenamefont{Jones et~al.}(2003)\citenamefont{Jones, Vale, Sahagun,
  Hall, and Hinds}}]{jonessurf}
\bibinfo{author}{\bibfnamefont{M.~P.~A.} \bibnamefont{Jones}},
  \bibinfo{author}{\bibfnamefont{C.~J.} \bibnamefont{Vale}},
  \bibinfo{author}{\bibfnamefont{D.}~\bibnamefont{Sahagun}},
  \bibinfo{author}{\bibfnamefont{B.~V.} \bibnamefont{Hall}}, \bibnamefont{and}
  \bibinfo{author}{\bibfnamefont{E.~A.} \bibnamefont{Hinds}},
  \bibinfo{journal}{Phys. Rev. Lett.} \textbf{\bibinfo{volume}{91}},
  \bibinfo{pages}{080401} (\bibinfo{year}{2003}).

\bibitem[{\citenamefont{Jo\textit{ et al.}}(2007{\natexlab{a}})}]{jointerf}
\bibinfo{author}{\bibfnamefont{G.~B.} \bibnamefont{Jo\textit{ et al.}}},
  \bibinfo{journal}{Phys. Rev. Lett.} \textbf{\bibinfo{volume}{98}},
  \bibinfo{pages}{180401} (\bibinfo{year}{2007}{\natexlab{a}}).

\bibitem[{\citenamefont{Wang\textit{ et al.}}(2005)}]{wanginterf}
\bibinfo{author}{\bibfnamefont{Y.-J.} \bibnamefont{Wang\textit{ et al.}}},
  \bibinfo{journal}{Phys. Rev. Lett.} \textbf{\bibinfo{volume}{94}},
  \bibinfo{pages}{090405} (\bibinfo{year}{2005}).

\bibitem[{\citenamefont{Schumm\textit{ et al.}}(2005)}]{schumminterf}
\bibinfo{author}{\bibfnamefont{T.}~\bibnamefont{Schumm\textit{ et al.}}},
  \bibinfo{journal}{Nature Physics} \textbf{\bibinfo{volume}{1}},
  \bibinfo{pages}{57} (\bibinfo{year}{2005}).

\bibitem[{\citenamefont{Jo\textit{ et al.}}(2007{\natexlab{b}})}]{jointerf2}
\bibinfo{author}{\bibfnamefont{G.~B.} \bibnamefont{Jo\textit{ et al.}}},
  \bibinfo{journal}{Phys. Rev. Lett.} \textbf{\bibinfo{volume}{99}},
  \bibinfo{pages}{240406} (\bibinfo{year}{2007}{\natexlab{b}}).

\bibitem[{\citenamefont{Hall et~al.}(2007)\citenamefont{Hall, Whitlock,
  Anderson, Hannaford, and Sidorov}}]{halldwell}
\bibinfo{author}{\bibfnamefont{B.~V.} \bibnamefont{Hall}},
  \bibinfo{author}{\bibfnamefont{S.}~\bibnamefont{Whitlock}},
  \bibinfo{author}{\bibfnamefont{R.}~\bibnamefont{Anderson}},
  \bibinfo{author}{\bibfnamefont{P.}~\bibnamefont{Hannaford}},
  \bibnamefont{and} \bibinfo{author}{\bibfnamefont{A.~I.}
  \bibnamefont{Sidorov}}, \bibinfo{journal}{Phys. Rev. Lett.}
  \textbf{\bibinfo{volume}{98}}, \bibinfo{pages}{030402}
  (\bibinfo{year}{2007}).

\bibitem[{\citenamefont{Scott et~al.}(2008)\citenamefont{Scott, Judd, and
  Fromhold}}]{scottjuddinterf}
\bibinfo{author}{\bibfnamefont{R.~G.} \bibnamefont{Scott}},
  \bibinfo{author}{\bibfnamefont{T.~E.} \bibnamefont{Judd}}, \bibnamefont{and}
  \bibinfo{author}{\bibfnamefont{T.~M.} \bibnamefont{Fromhold}},
  \bibinfo{journal}{Phys. Rev. Lett.} \textbf{\bibinfo{volume}{100}},
  \bibinfo{pages}{100402} (\bibinfo{year}{2008}).

\bibitem[{foo({\natexlab{a}})}]{footsurf}
\bibinfo{note}{Ref. \cite{andreas1} has further details of the potential. We
  mimic atom losses close to the surface by using an imaginary potential, which
  starts at $y=23\:\mu$m and increases linearly with increasing $y$.}

\bibitem[{\citenamefont{Scott et~al.}(2005)\citenamefont{Scott, Martin,
  Fromhold, and Sheard}}]{scott}
\bibinfo{author}{\bibfnamefont{R.~G.} \bibnamefont{Scott}},
  \bibinfo{author}{\bibfnamefont{A.~M.} \bibnamefont{Martin}},
  \bibinfo{author}{\bibfnamefont{T.~M.} \bibnamefont{Fromhold}},
  \bibnamefont{and} \bibinfo{author}{\bibfnamefont{F.~W.}
  \bibnamefont{Sheard}}, \bibinfo{journal}{Phys. Rev. Lett}
  \textbf{\bibinfo{volume}{95}}, \bibinfo{pages}{073201}
  (\bibinfo{year}{2005}).

\bibitem[{\citenamefont{Steel\textit{ et al.}}(1998)}]{steelTW}
\bibinfo{author}{\bibfnamefont{M.~J.} \bibnamefont{Steel\textit{ et al.}}},
  \bibinfo{journal}{Phys. Rev. A} \textbf{\bibinfo{volume}{58}},
  \bibinfo{pages}{4824} (\bibinfo{year}{1998}).

\bibitem[{foo({\natexlab{b}})}]{footres}
\bibinfo{note}{The shortest length scale that we must resolve in the system is
  the minimum healing length ($\sim 0.3$\:$\mu$m) and the shortest timescale is
  the correlation time.}

\bibitem[{foo({\natexlab{c}})}]{footimp}
\bibinfo{note}{The phase imprint is asymmetric because the co-ordinate origin
  is positioned at the mid-point of two adjacent grating wires.}

\bibitem[{\citenamefont{Scott et~al.}(2003)\citenamefont{Scott, Martin,
  Fromhold, Bujkiewicz, Sheard, and Leadbeater}}]{scottol}
\bibinfo{author}{\bibfnamefont{R.~G.} \bibnamefont{Scott}},
  \bibinfo{author}{\bibfnamefont{A.~M.} \bibnamefont{Martin}},
  \bibinfo{author}{\bibfnamefont{T.~M.} \bibnamefont{Fromhold}},
  \bibinfo{author}{\bibfnamefont{S.}~\bibnamefont{Bujkiewicz}},
  \bibinfo{author}{\bibfnamefont{F.~W.} \bibnamefont{Sheard}},
  \bibnamefont{and}
  \bibinfo{author}{\bibfnamefont{M.}~\bibnamefont{Leadbeater}},
  \bibinfo{journal}{Phys. Rev. Lett} \textbf{\bibinfo{volume}{90}},
  \bibinfo{pages}{110404} (\bibinfo{year}{2003}).

\bibitem[{\citenamefont{Scott et~al.}(2004)\citenamefont{Scott, Martin,
  Bujkiewicz, Fromhold, Malossi, Morsch, Cristiani, and Arimondo}}]{scottari}
\bibinfo{author}{\bibfnamefont{R.~G.} \bibnamefont{Scott}},
  \bibinfo{author}{\bibfnamefont{A.~M.} \bibnamefont{Martin}},
  \bibinfo{author}{\bibfnamefont{S.}~\bibnamefont{Bujkiewicz}},
  \bibinfo{author}{\bibfnamefont{T.~M.} \bibnamefont{Fromhold}},
  \bibinfo{author}{\bibfnamefont{N.}~\bibnamefont{Malossi}},
  \bibinfo{author}{\bibfnamefont{O.}~\bibnamefont{Morsch}},
  \bibinfo{author}{\bibfnamefont{M.}~\bibnamefont{Cristiani}},
  \bibnamefont{and} \bibinfo{author}{\bibfnamefont{E.}~\bibnamefont{Arimondo}},
  \bibinfo{journal}{Phys. Rev. A} \textbf{\bibinfo{volume}{69}},
  \bibinfo{pages}{033605} (\bibinfo{year}{2004}).

\bibitem[{\citenamefont{Scott et~al.}(2006)\citenamefont{Scott, Hutchinson, and
  Gardiner}}]{scottnoise}
\bibinfo{author}{\bibfnamefont{R.~G.} \bibnamefont{Scott}},
  \bibinfo{author}{\bibfnamefont{D.~A.~W.} \bibnamefont{Hutchinson}},
  \bibnamefont{and} \bibinfo{author}{\bibfnamefont{C.~W.}
  \bibnamefont{Gardiner}}, \bibinfo{journal}{Phys. Rev. A}
  \textbf{\bibinfo{volume}{74}}, \bibinfo{pages}{053605}
  (\bibinfo{year}{2006}).

\bibitem[{foo({\natexlab{d}})}]{foottc}
\bibinfo{note}{The correlation time is the BEC's healing length divided by the
  maximal speed of sound.}

\bibitem[{foo({\natexlab{e}})}]{foot3rdorder}
\bibinfo{note}{In Figs.\ \ref{fig:diffpopns}(g), \ref{fig:diffpopns}(h),
  \ref{fig:diffpopns}(j) and \ref{fig:diffpopns}(k), the modulations have a
  range of wavelengths with the shortest being due to a small contribution from
  the 3rd diffraction order.}

\bibitem[{\citenamefont{Wimberger\textit{ et al.}}(2005)}]{wimbergerOL}
\bibinfo{author}{\bibfnamefont{S.}~\bibnamefont{Wimberger\textit{ et al.}}},
  \bibinfo{journal}{Phys. Rev. A} \textbf{\bibinfo{volume}{72}},
  \bibinfo{pages}{063610} (\bibinfo{year}{2005}).

\end{thebibliography}

\end{document}